\documentclass{tlp} 

\usepackage{latexsym}
\usepackage{amssymb}
\usepackage{xspace}
\usepackage{url}
\usepackage{graphicx}
\usepackage{epsfig}
\usepackage{color}
\usepackage{alltt}

\newtheorem{example}{Example} 

\begin{document}
\bibliographystyle{acmtrans}

\long\def\comment#1{}

\newcommand{\nop}[1]{}

\def\pInst{\p_1}
\def\pASa{\p_2}
\def\pASb{\p_3}
\def\iASa{I_1}
\def\iASb{I_2}

\newcommand{\posbody}[1]{\ensuremath{B^{+}(#1)}}
\newcommand{\negbody}[1]{\ensuremath{B^{-}(#1)}}
\newcommand{\body}[1]{\ensuremath{B(#1)}}
\newcommand{\head}[1]{\ensuremath{H(#1)}}
\newcommand{\multi}[1]{\ensuremath{\overline{2}^{#1}}}
\newcommand{\natnum}{\ensuremath{\mathbb{N}}}
\newcommand{\posnatnum}{\ensuremath{\mathbb{N^+}}}
\newcommand{\intnum}{\ensuremath{\mathbb{I}}}

\newcommand{\occ }[1]{\ensuremath{{\cal E}_{occ}(#1)}}
\newcommand{\ground}[1]{\ensuremath{Ground(#1)}}

\newcommand{\tuple}[1]{\langle#1\rangle}
\newcommand{\lparse}{\textsc{lparse}}

\newcommand{\nafsymbol}{\ensuremath{\mathtt{not}}\xspace}
\newenvironment{dlvcode}
{\begin{displaymath}\begin{array}{l}}
{\end{array}\end{displaymath}}
\newcommand{\dlvstyle}[1]{\mbox{\small $#1$}}


\newcommand{\code}[1]{\ensuremath{#1}\xspace}

\newcommand{\OrSymbol}{\ensuremath{\vee}}
\newcommand{\OrhSymbol}{\ensuremath{\vee_h}}
\newcommand{\Orh}{\ensuremath{\,\OrhSymbol\,}\xspace}

\newcommand{\uneq}{\ensuremath{<>}}

\newcommand{\BPp}[1]{\ensuremath{B_{\p_{#1}}}}
\newcommand{\UPp}[1]{\ensuremath{U_{\p_{#1}}}}
\newcommand{\GPp}[1]{\ensuremath{Ground(\p_{#1})}}

\newcommand{\HR}{\ensuremath{H(\R)}}
\newcommand{\BR}{\ensuremath{B(\R)}}
\newcommand{\BpR}{\ensuremath{B^+(\R)}}
\newcommand{\BnR}{\ensuremath{B^-(\R)}}



\newcommand{\TODO}[1]{{\bf\large TODO:} {\em #1 }}

\newcommand{\dlv}{{\sc DLV}\xspace}
\newcommand{\ontodlp}{{OntoDLP}\xspace}
\newcommand{\ontodlv}{{OntoDLV}\xspace}

\newcommand{\derives}{\ensuremath{\mathtt{\ :\!\!-\ }}}
\newcommand{\Or}{\ensuremath{\vee}}
\newcommand{\Comma}{\,,\ }
\newcommand{\naf}{\ensuremath{\mathtt{not}}\xspace}
\newcommand{\countagg}{\ensuremath{\mathtt{\# count}}}
\newcommand{\sumagg}{\ensuremath{\mathtt{\# sum}}}
\newcommand{\minagg}{\ensuremath{\mathtt{\# min}}}
\newcommand{\maxagg}{\ensuremath{\mathtt{\# max}}}
\newcommand{\timesagg}{\ensuremath{\mathtt{\# times}}}
\newcommand{\anyagg}{\ensuremath{\mathtt{\# any}}}
\newcommand{\avgagg}{\ensuremath{\mathtt{\# avg}}}

\def\punto{\hspace*{\fill}{\rule[-0.5mm]{1.5mm}{3mm}}}

\newcommand{\GR}{\ensuremath{Ground(\R)}}
\newcommand{\R}{\ensuremath{r}}
\newcommand{\tneg}{\ensuremath{\neg}}

\newcommand{\p}{\ensuremath{{\cal P}}\xspace}
\newcommand{\GP}{\ensuremath{Ground(\p)}\xspace}
\newcommand{\BP}{\ensuremath{B_{\p}}\xspace}
\newcommand{\UP}{\ensuremath{U_{\p}}\xspace}

\newcommand{\NP}{{\rm NP}\xspace}
\newcommand{\SigmaP}[1]{{\Sigma}_{#1}^{P}}
\newcommand{\CONP}{\textrm{co-NP}\xspace}
\newcommand{\PiP}[1]{{\Pi}_{#1}^{P}}

\newcommand{\iec}[0]{i.e.,\ }
\newcommand{\egc}[0]{e.g.,\ }

\newcommand{\mytt}[1]{{\small\texttt{#1}}\xspace}

\title[Team-building with ASP in Gioia-Tauro Seaport]{Team-building with Answer Set Programming in the Gioia-Tauro Seaport}

\author[F. Ricca et. al]
{ F. Ricca$^1$, G. Grasso$^{1,3}$, M. Alviano$^1$, M. Manna$^1$, V. Lio$^2$, S. Iiritano$^2$, N. Leone$^1$\\
$^1$Dipartimento di Matematica, Universit{\`a} della Calabria, 87030 Rende, Italy\\
$^2$Exeura s.r.l., Via Pedro Alvares Cabrai - C.da Lecco 87036 Rende (CS), Italy\\
$^3$Computing Laboratory, University of Oxford, UK\\
E-mail: \{ricca,grasso,alviano,manna,leone\}@mat.unical.it, \\
\ \ \,\{vincenzino.lio,salvatore.iiritano\}@exeura.it
}

\pagerange{\pageref{firstpage}--\pageref{lastpage}}
\volume{\textbf{10} (3):}
\jdate{March 2002}
\setcounter{page}{1}
\pubyear{2002}

\submitted{24 July 2010}
\revised{4 Jan 2011}
\accepted{16 Jan 2011}

\maketitle

\label{firstpage}

\begin{abstract}
The seaport of Gioia Tauro is the largest transshipment terminal of the Mediterranean coast.
A crucial management task for the companies operating in the seaport is
team building: the problem of properly allocating the available personnel for serving the incoming ships.
Teams have to be carefully arranged in order to meet several constraints, such as
allocation of the employees with the appropriate skills, fair distribution of the working
load, and turnover of the heavy/dangerous roles.
This makes team building a hard and expensive task requiring several hours per day of manual preparation.

In this paper we present a system based on Answer Set Programming (ASP)
for the automatic generation of  the teams of employees in the seaport of Gioia Tauro.
The system is currently exploited in the Gioia Tauro seaport by ICO BLG,
a company specialized in automobile logistics.
\end{abstract}

\begin{keywords}
Answer Set Programming, Declarative Problem Solving, Knowledge Management,
Workforce Management, Artificial Intelligence.
\end{keywords}


\section{Introduction}\label{sec:intro}
In the last few years, the need for knowledge-based technologies has emerged in
several application areas.  Industries are now looking for \textit{semantic}
instruments for knowledge management 
enabling complex domain-knowledge modeling and real-world problem solving.
We report on a recent successful industrial application we developed in the area of Knowledge Management (KM), which is based on Answer Set Programming (ASP)~\cite{gelf-lifs-91}.
ASP is a fully-declarative language for Knowledge Representation and Reasoning (KRR),
which has been developed in the field of logic programming and nonmonotonic reasoning.
ASP has been already exploited for solving complex knowledge-based problems
arising in Artificial Intelligence~\cite{bara-gelf-2000,bald-etal-01,bara-uyan-2001,frie-08-techrep,fran-etal-2001,gebs-etal-2007-lpnmr-competition,noge-etal-2001}
as well as in Information Integration~\cite{leon-etal-2005},
and other areas of Knowledge Management~\cite{bara-2002,bard-95,grass-etal-09-apps-lpnmr}.
In recent years, the high knowledge-modeling power of this language
has been made available to users, who need to represent and manipulate complex knowledge,
by the development of some efficient ASP systems, such as \dlv~\cite{leon-etal-2002-dlv}.

In this paper, we present a new system, based on the ASP solver \dlv,
which is currently exploited by the ICO BLG company
in the international seaport of Gioia Tauro. The system allows for
the automatic generation of teams of employees.
Roughly, the problem we deal with is a {\em workforce management} problem~\cite{nave-etal-2007,erns-etal,tien-etal,lau,yang},
which amounts to computing a suitable
allocation of the available personnel of the seaport
such that cargo ships mooring in the port
are properly handled.
To accomplish this task several constraints have to be satisfied.
An appropriate number of employees, providing several different skills,
is required depending on the size and the load of cargo ships.
Moreover, the way an employee is selected and the specific role she will play in the team
(each employee is able to cover several roles according to her skills)
are subject to many conditions (\egc fair distribution of the working
load, turnover of the heavy/dangerous roles, etc.).
Up to now, the management of ICO BLG has been obliged to dedicate several hours
per day to the difficult and delicate task of allocating teams of employees
to incoming ships. Indeed, a bad allocation might cause delays or violations
of the contract with shipping companies, with
consequent pecuniary sanctions.
Moreover, an unfair distribution of the workload can lead to problems
with human resources management, ranging from: (i) the unavailability of crucial human resources
-when highly-skilled employees are not allocated in the best way-
(ii) to the loss of employee happiness  -resulting from an unfair distribution of heavy/dangerous roles-
that can directly affect production and performance.
Thus, team building is definitively a crucial and difficult management task for ICO BLG,
for which we developed an effective software solution.

The system described in this paper can either build new teams or automatically complete a partial allocation
when the user manually pre-assigns some key employees.
Moreover, it is able to verify whether a manually-composed team satisfies all the requirements, or not.
It might be the case that there is no team satisfying all requirements, or
that a team violates some condition due to user pre-assignments. In these cases,
the system provides a suitable explanation of violations
for allowing the user to take corrective actions.
Further, the system can also be used by the managers of the company for performing simulations in order to estimate important financial aspects. For instance, by computing ahead the whole shift scheduling for the next month, it is possible to evaluate how much overtime will be needed and
allow/deny leave requests.
Notably, in this application, the domain knowledge is declaratively modeled by exploiting ASP
and implemented by using the ASP system \dlv.
Desired allocations are computed by means of a set of suitably defined ASP programs.

The contribution of this paper is twofold:
firstly, we report on a successful industrially-developed KM system
which is currently employed in the important port of Gioia Tauro
(the system is commercially-distributed by EXEURA s.r.l.,
a spin-off company of the University of Calabria);
secondly, we describe how ASP can be profitably exploited for solving
a specific KM problem. The work
represents a practical demonstration  that ASP is well-suited
for developing effective real-world KM systems.

\medskip

In the literature there is a number of systems and methods for solving
various variants of the workforce management problem~\cite{nave-etal-2007,erns-etal,tien-etal,lau,yang}.
The available solutions can be classified in two main categories:
(i) ad-hoc algorithms, and
(ii) generic methods.
The first category has the drawback of producing solutions that
are very specific which cannot be easily adapted for solving
a different variants of the same problem; in practice they are employed
for solving problems with unique features
(see e.g., \citeNP{burke-soub,hult-card-97}).
Conversely, our approach belongs  to the second category where, historically,
both Operational Research and Artificial Intelligence methods
(such as Constraint Programming, Fuzzy logic, Genetic Algorithms, and Local Search)
were employed (see e.g., \citeNP{nave-etal-2007,erns-etal,tien-etal,lau,yang,gaud-ceru,lesa-etal,eitz-etal,dech-2004,gres-etal,bech-etal,alfa-2002,billi,ming-etal,aick-dows,wren,chiu-etal,ross-2000,sale-yako-2007,alb-chic-2007}).%
\footnote{For an in-depth comparison/characterization of solved problems
and adopted methods we refer the reader to the following
survey papers~\cite{nave-etal-2007,erns-etal,tien-etal,lau,yang}.}
%
The solutions presented in this paper also demonstrate that  ASP can be exploited
on a par with other AI techniques
for developing effective workforce management systems.
The distinctive features of our approach are:
\begin{itemize}

\item the distinction of individual employees%
\footnote{A feature missing in most of the existent approaches, as observed by~\citeN{nave-etal-2007}.}
by taking into account (i) specific employee-skills, (ii) current allocation history, and (iii)
temporary unavailability (owing to illness or special permission);

\item the satisfaction of practically-relevant cooperative work requirements, like the turnover of heavy/dangerous roles; and ``double shift'' allocations (where the same workers have to be employed in two adjacent shifts while respecting constraints);

\item system interactivity: the user can manually modify the provided solutions, since the system is able to deal with partial allocations and provide explanations when constraints are violated by manual modification;

\item high flexibility and maintainability: the system exploits a flexible, and easy-to-maintain core logic;

\item user friendly graphical interface that makes the use of the underlying technologies transparent to the user.

\end{itemize}

An important lesson we learned in this project is that the high expressiveness
of ASP language, which allows us to obtain an executable specification,
is an important advantage that can lead to prefer ASP over other approaches
in applications of this kind.
Indeed, the expressiveness of ASP and its purely declarative nature allowed us
to refine and fine-tune both problem specifications and encodings together,
while interacting with the stakeholders of the port.
We could easily simulate specification changes and immediately show their
effects to the customers.
The possibility of modifying (simply by editing a text file)
a complex reasoning task (e.g., by adding new constraints or by changing existing ones) in a few minutes
and testing
it ``on-site'' together with the customer is a great advantage of our approach,
it was highly appreciated by our customers, and constituted a key factor
for the success of our application.
We believe that such an advantage could be exploited in several contexts.

The result is a system that perfectly matches the requirements of the customer by allowing
both fast planning and fair employee allocation, while respecting the constraints.
The management of ICO BLG can now produce an admissible personnel allocation
in a few minutes: they spend less time and obtain a more effective
planning of the resources.

\medskip

The remainder of this paper is organized as follows: in the next section we give an overview of ASP;
Section~\ref{sec:domain} provides a specification of the team building problem;
while in Section~\ref{sec:encoding} we give an ASP encoding of the problem;
Section~\ref{sec:system} describes the architecture of the system and its functionalities;
Section~\ref{sec:exp} reports on the performances of the system on real-world data;
Section~\ref{sec:conclusion} concludes the paper.

\section{Answer Set Programming}\label{sec:language}

In this section we recall syntax and semantics of Answer Set Programming (ASP).
More specifically, we present an enriched language allowing for using set-oriented functions, also known as aggregate functions.
For further background on the standard ASP language we refer to \citeN{bara-2002} and \citeN{gelf-leon-02}.
For details on ASP with aggregate functions, instead, we refer to \citeN{fabe-etal-2004-jelia} and \citeN{lee-meng-2009-lpnmr}.

\subsection{Syntax}

We assume sets of \emph{variables}, \emph{constants}, and \emph{predicates} to be given.
Similar to Prolog, we assume variables to be strings starting with uppercase letters and constants to be non-negative integers or strings starting with lowercase letters.
Predicates are strings starting with lowercase letters.
An \emph{arity} (non-negative integer) is associated with each predicate.
Moreover, the language allows for using built-in predicates (i.e., predicates with a fixed meaning) for the common arithmetic operations (i.e., $=$, $\le$, $<$, $>$, $\ge$, $+$, etc.; usually written in infix notation).

\paragraph{Standard Atoms.}
A {\em term} is either a variable or a constant.
A {\em standard atom} is an expression \mytt{p(t$_1$,...,t$_k$)}, where \mytt{p} is a {\em predicate} of arity $k$ and \mytt{t$_1$,...,t$_k$} are terms.
If \mytt{t$_1$,...,t$_k$} are constants, \mytt{p(t$_1$,...,t$_k$)} is a {\em ground standard atom}.

\paragraph{Set Terms.}
A {\em set term} is either a symbolic set or a ground set.
A {\em symbolic set} is a pair $\{\mathit{Terms}:\ \mathit{Conj}\}$, where $\mathit{Terms}$ is a list of terms (variables or constants) and $\mathit{Conj}$ is a conjunction of standard atoms.
Intuitively, a set term \mytt{X: a(X,Y), b(Y)} stands for the set of \mytt{X}-values making the conjunction \mytt{a(X,Y), b(Y)} true, namely $\{\mytt{X} \mid \exists\,\mytt{Y}\ \mathit{s.t.} \ \mytt{a(X,Y), b(Y)} \ \mathit{is\ true}\}$.
A {\em ground set} is a set of pairs of the form $\tuple{\mathit{consts}\! :\!\mathit{conj}}$, where $\mathit{consts}$ is a list of constants and $\mathit{conj}$ is a conjunction of ground standard atoms.

\paragraph{Aggregate Functions.}
An {\em aggregate function} is of the form $f(S)$, where $S$ is a set term and $f$ is an {\em aggregate function symbol}.
Intuitively, an aggregate function can be thought of as a (possibly partial) function, mapping multisets of constants to constants.
Hereinafter, we will adopt the notation of the \dlv system~\cite{leon-etal-2002-dlv} for representing aggregate functions:
\begin{itemize}
\item \mytt{\#min \ \ } (minimal term, undefined for empty set);
\item \mytt{\#max \ \ } (maximal term, undefined for empty set);
\item \mytt{\#count\ } (number of terms);
\item \mytt{\#sum \ \ } (sum of integers).
\end{itemize}

\paragraph{Aggregate Atoms and Literals.}
An {\em aggregate atom} is a structure of the form $f(S) \odot T$, where $f(S)$ is an aggregate function, $\odot\ \in \{ <,\ \leq, >, \geq, =, \neq \}$ is a comparison operator, and $T$ is a term (a variable or a constant).
If $T$ is a constant and $S$ is a ground set term, $f(S) \odot T$ is a {\em ground aggregate atom}.
A {\em literal} is either
(i) a standard atom, or
(ii) a standard atom preceded by the {\em negation as failure} symbol \mytt{not}, or
(iii) an aggregate atom.

\paragraph{Programs.}
A {\em rule} \R{} is a construct of the form
\begin{dlvcode}
\alpha_1\ \Or\ \cdots\ \Or\ \alpha_n\ \derives\ \ell_1,\ \ldots,\ \ell_m.
\end{dlvcode}%
\noindent where $\alpha_1,\ \ldots,\ \alpha_n$ are standard atoms, $\ell_1,\ \ldots,\ \ell_m$ are literals, $n \geq 1$ and $m \geq 0$.
The disjunction $\alpha_1\ \Or\ \cdots\ \Or\ \alpha_n$ is referred to as the {\em head} of \R{}, while the conjunction $\ell_1,\ \ldots,\ \ell_m$ is the {\em body} of \R{}.
We denote by $\head{\R}$ the set of head atoms and by $\body{\R}$ the set of body literals.
If $\head{\R}$ and all the literals in $\body{\R}$ are ground, $\R$ is a {\em ground rule}.
A ground rule $\R$ is a {\em fact} if both $\body{\R}=\emptyset$ and $|\HR| = 1$.
A {\em program} $\p$ is a set of rules; if all rules in $\p$ are ground, $\p$ is a ground program.

\paragraph{Safety.}
A {\em local} variable of a rule $\R$ is a variable appearing solely in set terms of $\R$;
a variable of $\R$ that is not local is {\em global}.
A rule $\R$ is {\em safe} if both the following conditions hold:
(i) each global variable appears in some positive standard literal of $\body{\R}$;
(ii) each local variable appearing in a set term $\{\mathit{Terms}:\ \mathit{Conj}\}$ also appears in $\mathit{Conj}$.
A program is safe if all its rules are safe.
In the following, we assume that programs are safe.

\begin{example}\label{ex:safety}\rm
Consider the following rules:
\begin{alltt}\small
 c(X) :- q(X,Y,V), #max\{Z: a(Z), b(Z,V)\} \(>\) Y.
 c(X) :- q(X,Y,V), #sum\{Z: a(X), b(X,S)\} \(>\) Y.
 c(X) :- q(X,Y,V), #min\{Z: a(Z), b(Z,V)\} \(>\) T.
\end{alltt}\normalsize
The first rule is safe, while the second is not because the local variable \mytt{Z} violates condition (ii).
The third rule is not safe because the global variable \mytt{T} violates condition (i).
\punto
\end{example}

\subsection{Answer Set Semantics}
\label{sec:semantics}

Semantics of an ASP program is given by the set of its stable models, defined in this section.

\paragraph{Universe and Base.}
Given an ASP program \p, the {\em universe} of $\p$, denoted by \UP, is the set of constants appearing in \p.%
\footnote{%
If a program $\p$ has no constants, an arbitrary constant symbols $\xi$ is introduced.
}
The {\em base} of $\p$, denoted by \BP, is the set of standard atoms constructible from predicates of \p\ with constants in \UP.

\paragraph{Instantiation.}
A {\em substitution} is a mapping from a set of variables to $\UP$.
Given a substitution $\sigma$ and an ASP object $obj$  (atom, rule, set term, etc.), we denote by $obj\,\sigma$ the object obtained by replacing each variable \mytt{X} in $obj$ by $\sigma(\mytt{X})$.
A substitution from the set of global variables of a rule $\R$ (to $\UP$) is a {\em global substitution} for $\R$;
a substitution from the set of local variables of a set term $S$ (to $\UP$) is a {\em local substitution} for $S$.
Given a set term without global variables $S = \{\mathit{Terms}\! :\! \mathit{Conj}\}$, the \emph{instantiation of $S$} is the following ground set term:
$$inst(S) = \{ \tuple{\mathit{Terms}\,\sigma:\ \mathit{Conj}\,\sigma} \mid \sigma \mbox{ is a local substitution for } S\}.$$
A {\em ground instance} of a rule $\R$ is obtained in two steps:
first, a global substitution $\sigma$ for $\R$ is applied;
then, every set term $S$ in $\R\sigma$ is replaced by its instantiation $inst(S)$.
The instantiation \GP\ of a program \p\ is the set of instances of all rules in \p.

\begin{example}\label{ex:instantiation}
Consider the following program $\pInst$:
\begin{alltt}\small
 a(1) v b(2,2).
 a(2) v b(2,1).
 c(X) :- a(X), #sum\{Y: b(X,Y)\} \(\geq\) 2.
\end{alltt}\normalsize
The instantiation $\ground{\pInst}$ of $\pInst$ is the following program:
\begin{alltt}\small
 a(1) v b(2,2).
 a(2) v b(2,1).
 c(1) :- a(1), #sum\{\(\langle\)1: b(1,1)\(\rangle\), \(\langle\)2: b(1,2)\(\rangle\)\} \(\geq\) 2.
 c(2) :- a(2), #sum\{\(\langle\)1: b(2,1)\(\rangle\), \(\langle\)2: b(2,2)\(\rangle\)\} \(\geq\) 2. \punto
\end{alltt}\normalsize
\end{example}

\paragraph{Interpretations.}
An {\em interpretation} $I$ for an ASP program \p\ is a subset of $\BP$.
A standard ground atom $\alpha$ is true with respect to $I$ if $\alpha \in I$;
otherwise, $\alpha$ is false with respect to $I$.
A negative standard ground literal \mytt{not} $\alpha$ is true with respect to $I$ if $\alpha \not\in I$;
otherwise, \mytt{not} $\alpha$ is true with respect to $I$.

An interpretation $I$ also provides a meaning to set terms, aggregate functions and aggregate literals, namely a multiset, a value, and a truth value, respectively.
The evaluation $I(S)$ of a set term $S$ with respect to $I$ is the multiset $I(S)$ defined as follows:
Let $S^I=\{\tuple{t_1,...,t_k}\mid \tuple{t_1,...,t_k\!:\!\mathit{Conj}}\!\in\! S$ and all the atoms in $Conj$ are true with respect to $I\}$;
then, $I(S)$ is the multiset obtained as the projection of the tuples of $S_I$ on their first constant, that is, $I(S)=[t_1 \mid \tuple{t_1,...,t_n} \in S^I ]$.
The evaluation $I(f(S))$ of an aggregate function $f(S)$ with respect to $I$ is the result of the application of $f$ on $I(S)$.
If the multiset $I(S)$ is not in the domain of $f$, $I(f(S))= \bot$ (where $\bot$ is a fixed symbol not occurring in \p).
A ground aggregate atom $f(S) \odot k$ is true with respect to $I$ if both
$I(f(S))\neq \bot$ and
$I(f(S)) \odot k$ hold;
otherwise, $f(S) \odot k$ is false.

\paragraph{Models.}
Given an interpretation $I$, a rule $r$ is {\em satisfied} with respect to $I$ if some head atom is true with respect to\ $I$ whenever all body literals are true with respect to\ $I$.
An interpretation $M$ is a {\em model} of an ASP program $\p$ if all the rules $\R \in \GP$ are satisfied with respect to\ $M$.
A model $M$ for $\p$ is (subset) minimal if no model $N$ for $\p$ exists such that $N \subsetneq M$.%
\footnote{%
Note that an answer set $M$ of $\p$ is also a model of $\p$.
Indeed, $\GP^M \subseteq \GP$ and rules in $\GP \setminus \GP^M$ are satisfied with respect to $M$ by definition of reduct.
}

\paragraph{Answer Sets.}
We now recall the generalization of the Gelfond-Lifschitz transformation and answer sets for programs with aggregates from \citeN{fabe-etal-2004-jelia}:
Let $\p$ be a ground program, $I$ be an interpretation, and $\p^I$ denote the transformed program obtained from $\p$ by deleting all rules in which a body literal is false with respect to\ $I$.
An interpretation $M$ is an answer set of the program $\p$ if it is a minimal model of $\GP^M$.

\begin{example}\label{ex:simple-trans}
Consider two interpretations $\iASa =$ \mytt{\{a(0)\}} and $\iASb = $ \mytt{\{b(0)\}} for the next programs:
\begin{alltt}\small
 \(\pASa\) = \{a(0) :- #count\{X:\ b(X)\} \(>\) 0.\}
 \(\pASb\) = \{a(0) :- #count\{X:\ b(X)\} \(\leq\) 0.\}
\end{alltt}\normalsize
Hence, we obtain the following transformed programs:
\begin{alltt}\small
 \(\ground{\pASa}\sp{\iASa}\) \(=\) \(\emptyset\)
 \(\ground{\pASa}\sp{\iASb}\) \(=\) \(\ground{\pASa}\) \(=\) \{a(0) :- #count\{\(\langle\)0: b(0)\(\rangle\)\} \(>\) 0.\}
 \(\ground{\pASb}\sp{\iASa}\) \(=\) \(\ground{\pASb}\) \(=\) \{a(0) :- #count\{\(\langle\)0: b(0)\(\rangle\)\} \(\leq\) 0.\}
 \(\ground{\pASb}\sp{\iASb}\) \(=\) \(\emptyset\)
\end{alltt}\normalsize
Hence, neither $\iASa$ nor $\iASb$ are answer sets of $\pASa$, while $\iASa$ is an answer set of $\pASb$ because $\emptyset$ is not a model of $\ground{\pASb}^{\iASa}$.
On the other hand, $\iASb$ is not an answer set of $\pASb$ because $\emptyset$ is a model of $\ground{\pASb}^{\iASb}$.
\punto
\end{example}

Note that the syntax of the language does not explicitly allow rules without head atoms, called constraints.
However, constraints can be simulated by using a new symbol and negation.
For example, a constraint of the form $\derives\ \ell_1,\ \ldots,\ \ell_m$ can be replaced by a rule of the form $\mathit{co} \derives\ \ell_1,\ \ldots,\ \ell_m,\ \mytt{not}\ \mathit{co}$, where $\mathit{co}$ is a symbol not occurring in \p.

\section{Problem Motivation and Specification}\label{sec:domain}
The seaport of Gioia Tauro%
\footnote{See \url{http://www.portodigioiatauro.it}}
is the largest
transshipment terminal of the Mediterranean coast.
Most of the transported goods reach and leave this port by ship,
and only a minimal part of them (about 5\%) is transferred by train or truck.
Historically, container transshipment has been the main activity of the port
(related problems were the subject of extensive research; see e.g., \citeNP{vacca-etal-2007}),
while  recently Gioia Tauro has become also an automobile hub.
Automobile logistics is carried out by the ICO BLG company%
\footnote{ICO BLG is a subsidiary of the BLG Logistics Group --- \url{http://www.blg.de}}
on an area of  320,000 m$^2$ corresponding to a warehouse capacity for about 15,000 vehicles.
In the last few years, cargo traffic has increased impressively in Gioia Tauro:
every day several vessels of different capacities moor in the port carrying their load of new cars and trucks.
The vehicles they transport might have to be handled, warehoused, processed
(if technically necessary) and finally delivered to their final destination.
Depending on the specific processing needs, the load of a mooring ship might be subject
to different activities,  each of which requires the allocation of a number of specifically-skilled employees.
For instance, a ship might be partially or fully loaded/unloaded,
and transported vehicles identified, quality-checked and moved to/from the port yard.
Moreover, cars (possibly-damaged during transportation) might undergo an additional repairing
process. The port is, in fact, equipped with a centre able
to repair both minor damage to car bodywork and to perform
pre-delivery services such as polishing, washing, battery testing, etc.

The time required for processing ships is set up by contract.
For this reason, a critical management goal is  to  serve arriving boats promptly
because any delay might cause financial penalties to the company.
In order to accomplish that task, several teams of employees have
to be arranged and scheduled to  work simultaneously on these ships.
The selection of the employees and the assignment of roles is subject to many conditions.
Some constraints are imposed by the employees skills and availability,
other by contract (in general identified as legal conditions).
For example, each team usually includes at least one ``quality-checker'', and
enrolls a sufficient number of drivers for loading/unloading cars in time.
Moreover, each employee cannot work more than 36 hours (plus max 12 hours of overtime) per week
and, importantly, heavy/dangerous roles must be turned over.
Note that some tasks require that employees work in ``bad conditions'';
for instance, the employees performing operations in the hold (while cars are driven in/out
of a ship) are persistently exposed to a high concentration of pollution.
Hence, daily assignments to such heavy roles must be turned over among the personnel,
so that hard and easy tasks are shared among employees.
Moreover, a fair distribution of work has to be obtained by ensuring
that all the employees are enrolled for about the same number of hours in a week.
In this way, both the best possible working conditions for the employees
and the processing of all incoming boats are guaranteed.

Data regarding arriving/departing ships,
such as date and time, number and kind of transported vehicles,
are usually known by the managers (at least) one day in advance.
Given these data, the management easily identifies the team composition characteristics
needed for processing a given ship (i.e., how many employees are needed,
what skills the workers must have, how the working time is divided in shifts, etc.).
According to ICO BLG terminology, we say that a meta-plan is
a specification of the needed skills and number of employees for one shift.
(A meta-plan usually refers to a single ship.)
However, the subsequent task of assigning the available employees
to shifts and roles (remember that each employee is able to cover several roles according to her skills),
in such a way that all the requirements are satisfied,
is definitively the hardest part of the whole rostering process.

The ICO BLG managers were used to find out a ``hand-made'' solution.
Of course, this technique was far from being effective and efficient.
Indeed, arranging teams required many hours, often resulting in solutions not
fulfilling several constraints.
The impossibility of allocating teams to incoming ships might cause
delays and violations of the contract with employees or with shipping companies.
As a consequence, {\em team building is definitively a crucial management task} for ICO BLG.

In the following, all the requirements to be fulfilled are specified in more detail.

\subsection{Team Requirements and Available Personnel}
Information regarding mooring ships, such as arrival and departure time
as well as their processing needs (load/unload, size of the load, etc.), is known
one day in advance, and the rostering process is always carried out by considering one day at time.
Moreover, working time of a day is divided in shifts lasting 6-12 hours each.
Given this information, the  managers can easily produce
a specification of the resources needed during the working shifts for dealing with the incoming ships.
In particular,  the following information is specified: the shift data (date-time identifier and duration),
the set of skills to be covered, and the number of needed employees per skill.
This is the main input of our team building system because the goal
is to select the right workers among those which are available for
fulfilling resource requests.
However, the availability of a specific employee for a given shift depends on several conditions,
some of which are stated by the employees contract, some by other circumstances.

Concerning legal requirements, we already mentioned that an employee
cannot exceeds a certain amount of working hours (including overtime) per day and per week.
These values are given as input parameters within our system, together with the allocation
history recording the number of elapsed working hours for each employee.
A further condition imposed by the contract regards night-shifts.
It states that an employee who has worked during a night-shift
must have a mandatory rest time before being allocated once more.
In addition, an employee can be assigned to a single vessel in a given shift, but she can be moved to a different vessel in a subsequent shift.
An employee cannot change role in a shift, but she can play different roles in different shifts.
Finally, an employee is not \emph{available} if either she is in vacation, or
a specific management decision requires that she must not be a member of the team.

\subsection{Turnover and Fairness Requirements}

As previously pointed out, to obtain the best possible working conditions
for the employees, heavy/dangerous roles have to be turned over.
More in detail, a ``round-robin like'' turnover policy should be applied for avoiding that the same group of employees is mostly allocated to heavy/dangerous roles.
Moreover,
a fair distribution of the workload has to be ensured in such a way that
all the employees have to be allocated for about the same amount of weekly hours.
This requirement concerns the way the workload is spread over employees within a week.
As an example, suppose that the workload expected for a given week is relatively low.
In this case, the possible solutions range from the exploitation of the same few employees
(possibly up to their weekly limit of hours) and the distribution of the few shifts to
distinct available employees. Clearly, the last scenario is the fairest and, in general,
it should be avoided that a group of employees works less (or is unemployed) in a week,
whereas some other group is overloaded. Basically, the larger admissible gap
of worked hours among employees has to be maintained below a threshold
(which can be set as a parameter in our system) of usually 8 hours.

\subsection{Crucial Roles}

The employees of ICO BLG, in general, can play different roles in a team,
depending on their skills. For instance, the same employee can be enrolled in a team
as a driver (her role is to drive cars in/out of the boats) or as a quality-checker,
in case she is skilled on both activities.
Clearly, there are a few employees skilled in critical roles
(such as quality-checkers),
while almost all employees are skilled for easy roles (e.g., all employees are drivers).
If employees skilled in critical roles are not employed carefully, it might be the case that
a team satisfying all the requirements cannot be built.
Indeed, if these few employees are assigned to other (more common) roles
during the first days of the week, it might happen that they become unavailable on the last days of the week
(e.g., they may have already reached 36 working hours). 
In order to avoid these disadvantageous situations, employees having crucial roles have to be preserved if possible.

\subsection{Double Shift}
A very specific requirement of our team building problem is the possibility of specifying {\em double shifts}.
So far, we have just considered (simple) shifts, where employees starts and
finish working performing the same role.
Instead, a double shift can be thought of as two consecutive (simple) shifts 
where the same employees are assigned to possibly different roles.%
\footnote{Note that this requirement cannot be enforced by considering single shifts only.}
More in detail, let $n_{1}$ and $n_{2}$ be the number of employees required
for two shift $s_{1}$ and $s_{2}$, respectively; a double shift has the following
property: if $s_{2}$ does not require a higher number of workers than $s_{1}$ (\iec{$n_{2} \leq n_{1}$}),
then only employees already working in $s_{1}$ have to be used in $s_{2}$;
otherwise ($n_{2} > n_{1}$),  all the employees working in $s_1$
have to be allocated in $s_{2}$ and the team has to be (possibly)  augmented
by taking additional employees.
Moreover, when moving from $s_{1}$ to $s_{2}$, roles must be turned over
with respect to the assignment of $s_{1}$.

\section{ASP-based Encoding}\label{sec:encoding}

This section describes the ASP program which solves
the team building problem specified in the previous section.
First, the input data is described; then, the ASP rules computing
teams of employees are described.

\subsection{Input Specification}
The input of the process is specified by means of the predicates described in this section.
The association among employees and skills is expressed by instances of \textit{hasSkill(employee, skill)}.
Instances of \textit{absent(employee, shift)} are used for storing information regarding
employees being not available on a given shift because
not present (e.g., for day off, for disciplinary decision, or for resting after a night-shift).
In addition, instances of \textit{manuallyExcluded(employee, shift)}
are used for modeling employees who do not have to be considered according
to some management decision
(to allow manual intervention on the allocation of specific employees).
Shifts are modeled by instances of \textit{shift(shiftId, duration)}, reporting the duration of each shift.
Moreover, the temporal sequence of shifts is represented by instances of \textit{previousShift(shift, shift$\,'$)}, where \textit{shift$\,'$} immediately precede \textit{shift}.
Resource requirements are represented
by instances of \textit{neededEmployees(shift, skill, n)},
reporting the number \textit{n} of needed employees for each shift and skill.

Statistics concerning past allocations are modeled by instances of
\textit{workedWeeklyHours(employee, hours)} and \textit{workedDailyHours(employee, hours)},
modeling worked hours for each employee up to the current allocation.
Similarly, instances of \textit{workedWeekOvertimeHours(employee, hours)} are used for modeling the count of overtime
for each employee up to the current allocation.
Instances of these predicates are produced
by querying an internal database that is not described here
to simplify the presentation.
Few additional parameters of the process are given by means of the following predicates:
an instance of \textit{dailyHours(hours)} for specifying the maximum numbers of working hours allowed per day;
an instance of \textit{weekHours(hours)} for setting the maximum numbers of working hours allowed per week;
an instance of \textit{weekOvertime(hours)} specifying how many overtime hours per week an employee is allowed to work.

The predicates described up to now constitute the input of the subprogram reported in Section~\ref{subsec:teamRec} and used for effectively produce all candidate solutions for the team building problem.
The input in enriched by predicates allowing for discarding candidate solutions violating preference requirements on heavy/dangerous (Section~\ref{subsec:turnover}) or crucial roles (Section~\ref{subsec:crucial}).
In particular, heavy/dangerous roles are specified by instances of \textit{heavyRole(skill)} and, for each employee and skill, the last allocation date of that employee on that skill is given by an instance of \textit{lastAllocation(employee, skill, lastTime)}.
Moreover, the maximum admissible difference of working hours among all the employees
is given by an instance of \textit{fairGap(hours)}.
Concerning crucial skills, they are identified by instances of \textit{crucialRole(skill)}.
Moreover, the input contains an instance of \textit{hasCrucial(employee, n)} for each employee, specifying the number \textit{n} of her crucial skills.

Finally, double shifts are modeled by instances of \textit{isDouble(shift$_1$, shift$_2$)},
where \textit{shift$_1$} and \textit{shift$_2$} are consecutive (simple) shifts,
and the number of employees required by \textit{shift$_1$} is less than or equal to the one
required by \textit{shift$_2$}.
These facts are used by the subprogram reported in Section~\ref{subsec:double}.

\subsection{Team Requirements and Available Personnel}\label{subsec:teamRec}

The following rule identifies
the set of employees who can be assigned to a specific role in each shift:
\begin{alltt}\small
 canBeAssigned(Em,Sh,Sk) :- hasSkill(Em,Sk), neededEmployees(Sh,Sk,_),
    not absent(Em,Sh), not manuallyExcluded(Em,Sh),
    not exceedTimeLimit(Em,Sh).
\end{alltt}\normalsize
In detail, an employee \mytt{Em} can be assigned to a shift \mytt{Sh} for a role \mytt{Sk}
if the following conditions are satisfied:
(i) she has the skill \mytt{Sk};
(ii) \mytt{Sh} needs employees with skill \mytt{Sk};
(iii) she is not absent
(iv) she is not excluded by some managers decision; and
(v) she does not exceed time limits.
Predicate \mytt{exceedTimeLimit} is intentionally defined as follows:

\begin{alltt}\small
 exceedTimeLimit(Em,Sh) :- shift(Sh,D),
    workedWeeklyHours(Em,Wh), weekHours(Hmax), D + Wh  \(>\) Hmax.

 exceedTimeLimit(Em,Sh) :- shift(Sh,D), dailyHours(Hmax),
    workedDailyHours(Em,Wh), D + Wh  \(>\) Hmax.

 exceedTimeLimit(Em,Sh) :- shift(Sh,D), weekOvertime(Hmax),
    workedWeekOvertimeHours(Em,Wh), D + Wh  \(>\) Hmax.
\end{alltt}\normalsize
In particular, the first rule can be read as follows:
``if a shift \mytt{Sh} has duration \mytt{D}
and, for a given employee \mytt{Em}, the sum of \mytt{D} with \mytt{Wh} (the hours worked by \mytt{Em} up to \mytt{Sh})
is greater than \mytt{Hmax} (the maximum allowed hours in a week),
then \mytt{Em} exceeds the time limit when associated to \mytt{Sh}''.
The remaining two rules can be interpreted in a similar way if daily worked hours and overtime are
considered instead.

Next, according to the guess\&check programming methodology
~\cite{eite-etal-2000c,leon-etal-2002-dlv}, the following disjunctive rule 
guesses the selection of employees that can be assigned to the shifts in the appropriate roles:
\begin{alltt}\small
 assign(Em,Sh,Sk) v nAssign(Em,Sh,Sk) :- canBeAssigned(Em,Sh,Sk).
\end{alltt}\normalsize
This rule can be read as follows: ``if an employee \mytt{Em} matches the skill requirements for a shift \mytt{Sh}, then
\mytt{Em} can be assigned to \mytt{Sh}, or not''.
Assignments violating team requirements are filtered out
by the following constraints:
\begin{alltt}\small
 :- neededEmployees(Sh,Sk,EmpNum), #count\{Em: assign(Em,Sh,Sk)\} \(\neq\) EmpNum.

 :- assign(Em,Sh,Sk\(\sb{1}\)), assign(Em,Sh,Sk\(\sb{2}\)), Sk\(\sb{1}\) \(\neq\) Sk\(\sb{2}\).

 :- assign(Em,Sh\(\sb{1}\),_), assign(Em,Sh\(\sb{2}\),_), Sh\(\sb{1}\) \(\neq\) Sh\(\sb{2}\). \hfill (*)
\end{alltt}\normalsize
In particular, the first constraint discards assignments with a wrong number of employees per required skill
(in other words, an assignment is discarded if the total count of employees assigned to a skill does not match requirements).
On the other hand, the second and third constraints avoid
that the same employee covers two roles or two shifts, respectively.%
\footnote{Recall that a single day is considered during an execution.}

\subsection{Turnover and Fairness Requirements}\label{subsec:turnover}

A careful assignment to heavy/dangerous roles is probably one of the most important
requirements for ensuring the best possible working conditions to employees.
We achieved this behavior by employing the following rule:
\begin{alltt}\small
 prefByTurnover(Em\(\sb{1}\),Em\(\sb{2}\),Sh,Sk) :- heavyRole(Sk),
    canBeAssigned(Em\(\sb{1}\),Sh,Sk), canBeAssigned(Em\(\sb{2}\),Sh,Sk),
    lastAllocation(Em\(\sb{1}\),Sk,Date\(\sb{1}\)), lastAllocation(Em\(\sb{2}\),Sk,Date\(\sb{2}\)), Date\(\sb{1}\) < Date\(\sb{2}\).
\end{alltt}\normalsize
According to this rule, if two employees \mytt{Em$_1$} and \mytt{Em$_2$} are both skilled
on a heavy role \mytt{Sk},
and \mytt{Em$_2$} has performed \mytt{Sk} more recently than \mytt{Em$_1$},
then \mytt{Em$_1$} is preferred to \mytt{Em$_2$}.
Therefore, for each shift and heavy role, predicate \mytt{prefByTurnover} states a priority ordering
among employees, and the following constraint implements rotation of roles
by forbidding assignments whenever turnover preferences are not respected:
\begin{alltt}\small
 :- prefByTurnover(Em\(\sb{1}\),Em\(\sb{2}\),Sh,Sk), assign(Em\(\sb{2}\),Sh,Sk), not assign(Em\(\sb{1}\),Sh,Sk).
\end{alltt}\normalsize
In particular, for a given shift, the assignment of an employee \mytt{Em$_2$} on a role \mytt{Sk} is forbidden
when there is an unassigned employee \mytt{Em$_1$} having higher priority than \mytt{Em$_2$} on \mytt{Sk}.
Clearly, this implies that admissible solutions are only those that actually ensure an effective turnover,
applying a ``round-robin'' policy among workers.
Following a similar approach, we can impose the fairness requirement for
spreading the workload among employees.
The next rule sets allocation priorities among available workers:
\begin{alltt}\small
 prefByFairness(Em\(\sb{1}\),Em\(\sb{2}\),Sh,Sk) :- fairGap(MaxGap),
    workedWeeklyHours(Em\(\sb{1}\),Wh\(\sb{1}\)), workedWeeklyHours(Em\(\sb{2}\),Wh\(\sb{2}\)),
    canBeAssigned(Em\(\sb{1}\),Sh,Sk), canBeAssigned(Em\(\sb{2}\),Sh,Sk), Wh\(\sb{1}\) + MaxGap < Wh\(\sb{2}\).
\end{alltt}\normalsize
In detail, if the gap on weekly worked hours between employees \mytt{Em$_2$} and \mytt{Em$_1$} exceeds
the maximum amount allowed, then \mytt{Em$_1$} has higher priority than \mytt{Em$_2$}.

Solutions violating fairness preferences
are discarded by the constraint
\begin{alltt}\small
 :- prefByFairness(Em\(\sb{1}\),Em\(\sb{2}\),Sh,Sk), assign(Em\(\sb{2}\),Sh,Sk), not assign(Em\(\sb{1}\),Sh,Sk).
\end{alltt}\normalsize

\subsection{Crucial Roles}\label{subsec:crucial}
In order to preserve crucial roles, the assignment of employees
qualified for few crucial skills has to be preferred.
Note that employees having many crucial skills may act as ``wildcards''
and should be employed with parsimony.
This aspect is encoded by the rule
\begin{alltt}\small
 prefByCrucial(Em\(\sb{1}\),Em\(\sb{2}\),Sh,Sk) :- hasCrucial(Em\(\sb{1}\),N\(\sb{1}\)), hasCrucial(Em\(\sb{2}\),N\(\sb{2}\)),
    canBeAssigned(Em\(\sb{1}\),Sh,Sk), canBeAssigned(Em\(\sb{2}\),Sh,Sk), N\(\sb{1}\) < N\(\sb{2}\).
\end{alltt}\normalsize
modeling a priority order that depends on the number of crucial skills of employees.
Finally, assignments not respecting preferences on crucial roles
are discarded by the following constraint:
\begin{alltt}\small
 :- prefByCrucial(Em\(\sb{1}\),Em\(\sb{2}\),Sh,Sk), assign(Em\(\sb{2}\),Sh,Sk), not assign(Em\(\sb{1}\),Sh,Sk).
\end{alltt}\normalsize
%
\subsection{Double Shifts}\label{subsec:double}
A double shift comprises two simple shifts in which an employee is possibly enrolled in different roles (one for each component).
However, constraint \mytt{(*)} in Section~\ref{subsec:teamRec} prevents to enroll an employee in different roles in a run.
Thus, for allowing the association of the same employees in a double shift, constraint \mytt{(*)} is modified as follows:
\begin{alltt}\small
 :- assign(Em,Sh\(\sb{1}\),_), assigned(Em,Sh\(\sb{2}\),_), Sh\(\sb{1}\) \(\neq\) Sh\(\sb{2}\), not isDouble(Sh\(\sb{1}\),Sh\(\sb{2}\)).
\end{alltt}\normalsize
Moreover, recall that an instance of \mytt{isDouble(Sh$_{\mathit{small}}$,Sh$_{\mathit{big}}$)} 
is in the input whenever \mytt{Sh$_{\mathit{small}}$} and \mytt{Sh$_{\mathit{big}}$} constitute a double shift, and
the {\em number of employees} required by \mytt{Sh$_{\mathit{small}}$} is less than or equal to the one
required by \mytt{Sh$_{\mathit{big}}$}.
Thus, the rules
\begin{alltt}\small
 :- isDouble(Sh\(\sb{\mathit{small}}\),Sh\(\sb{\mathit{big}}\)), assigned(Em,Sh\(\sb{\mathit{small}}\)), not assigned(Em,Sh\(\sb{\mathit{big}}\)).

 assigned(Em,Sh) :- assign(Em,Sh,_).
\end{alltt}\normalsize
enforce that all employees assigned to \mytt{Sh$_{\mathit{small}}$}
must be assigned to \mytt{Sh$_{\mathit{big}}$} as well.
Finally, for respecting bounds on weekly, daily, and overtime working hours, the following constraints are added:
\begin{alltt}\small
 :- isDouble(Sh\(\sb{1}\),Sh\(\sb{2}\)), assigned(Em,Sh\(\sb{1}\)), assigned(Em,Sh\(\sb{2}\))
    shift(Sh\(\sb{1}\),D\(\sb{1}\)), shift(Sh\(\sb{2}\),D\(\sb{2}\)), workedWeeklyHours(Em,Wh),
    weekHours(Hmax), Wh + D\(\sb{1}\) + D\(\sb{2}\)  \(>\) Hmax.

 :- isDouble(Sh\(\sb{1}\),Sh\(\sb{2}\)), assigned(Em,Sh\(\sb{1}\)), assigned(Em,Sh\(\sb{2}\))
    shift(Sh\(\sb{1}\),D\(\sb{1}\)), shift(Sh\(\sb{2}\),D\(\sb{2}\)), workedDailyHours(Em,Wh),
    weekHours(Hmax), Wh + D\(\sb{1}\) + D\(\sb{2}\)  \(>\) Hmax.

 :- isDouble(Sh\(\sb{1}\),Sh\(\sb{2}\)), assigned(Em,Sh\(\sb{1}\)), assigned(Em,Sh\(\sb{2}\))
    shift(Sh\(\sb{1}\),D\(\sb{1}\)), shift(Sh\(\sb{2}\),D\(\sb{2}\)), workedWeekOvertimeHours(Em,Wh),
    weekHours(Hmax), Wh + D\(\sb{1}\) + D\(\sb{2}\)  \(>\) Hmax.
\end{alltt}\normalsize
In particular, these constraints discard assignments in which an employee violates time constraints when enrolled in both components of a double shift.

\subsection{Dealing with Inconsistency}

The ASP program described so far provides admissible solutions of the team building problem
where all the requirements are satisfied.
It is easy to see that the combined action of different constraints
may cause situations where there is no admissible solution.
As an example, suppose we want to compose a team of two people covering two different roles,
where skill $s_{1}$ and $s_{2}$ are required. Suppose also that $s_{1}$ is a skill subject to turnover,
and $e_{1}$ is the only employee that can be assigned to $s_{1}$ for respecting this requirement;
moreover, all the employees but $e_{1}$ have worked for 20 hours so far,
while $e_{1}$ has just worked for 10 hours (note that such a situation is not unusual,
because $e_{1}$ may have worked during night in its last assignment).
According to the fairness requirement, no one should be preferred to $e_{1}$ for covering $s_{2}$.
This leads to a conflicting scenario where $e_{1}$ is the only possible choice for both $s_{1}$ and $s_{2}$,
but the same employee cannot cover two distinct roles in the same team.
Similar conflicting situations can occur in more complex settings where similar ``loops'' might happen.

Note that, in a real world scenario, it is often the case that
all the constraints cannot be simultaneously satisfied.
In order to deal with such a problem, together with ICO BLG managers, we devised a strategy applying
constraints in a prioritized way.
In particular, the managers of the company specified that the constraints
have to be applied in the following importance order:
(1) turnover, (2) fairness, and (3) crucial roles conditions
(where the application of the leftmost has higher precedence).
The idea is to avoid the application of more than one constraint on the same pair of employees.
For instance, if an employee \mytt{Em$_1$} is preferred to \mytt{Em$_2$} according to the turnover,
then fairness and crucial roles preferences must not be applied on the same workers.
Thus, even if \mytt{Em$_2$} would be preferable to \mytt{Em$_1$} because of the fairness constraint,
still \mytt{Em$_1$} should be possibly taken.
In order to implement this specification,
we devised a prioritized variant of the above-mentioned constraints:
\begin{alltt}\small
 :- prefByCrucial(Em\(\sb{1}\),Em\(\sb{2}\),Sh,Sk), assign(Em\(\sb{2}\),Sh,Sk),
    not assign(Em\(\sb{1}\),Sh,Sk), not prefByTurnover(Em\(\sb{1}\),Em\(\sb{2}\),Sh,Sk),
    not prefByFairness(Em\(\sb{1}\),Em\(\sb{2}\),Sh,Sk).

 :- prefByFairness(Em\(\sb{1}\),Em\(\sb{2}\),Sh,Sk), assign(Em\(\sb{2}\),Sh,Sk),
    not assign(Em\(\sb{1}\),Sh,Sk), not prefByTurnover(Em\(\sb{1}\),Em\(\sb{2}\),Sh,Sk).
\end{alltt}\normalsize
The first constraint applies crucial role priority among two employees
only if both turnover of roles and fairness requirements do not apply on the same employees.
In a similar way, the second constraint enforces fairness only if the turnover constraint
does not apply.
Note that the constraint implementing turnover requirement did not need any reformulation.
Indeed, they have the highest priority and, thus, have to be applied in any case.

Within our system, the encoding modified with constraint priority is run automatically
if no answer set is produced by the original encoding.
In this way, the system can find ``acceptable'' solutions
by ignoring constraints that are in conflict with ones of higher priority.
The system alerts the user when such a case occurs.

\subsection{Team Checking}
Upon an explicit customer request,
our system also allows the user to modify a computed team (partially or completely).
Whenever a team is manually modified,
the system checks whether it violates some constraints and informs the user.
The checking is carried out by running the same program above,
where the ``guessing'' rule is replaced by a set of facts for the ``assign''
predicate for specifying the team.

In case of inconsistency, one might be interested in finding the causes of the conflict.
To this end, a slightly modified version of the same program can be exploited for detecting violated constraints,
where specifically conceived atoms are introduced in the head of the constraints modeling preference conditions.
For example, the constraint requiring the turnover of heavy/dangerous roles is replaced by the
following rule:
\begin{alltt}\small
 violatedTurnover(Em\(\sb{1}\),Em\(\sb{2}\),Sh,Sk) :- prefByTurnover(Em\(\sb{1}\),Em\(\sb{2}\),Sh,Sk),
    assign(Em\(\sb{2}\),Sh,Sk), not assign(Em\(\sb{1}\),Sh,Sk).
\end{alltt}\normalsize
An instance of \mytt{violatedTurnover(Em$_1$,Em$_2$,Sh,Sk)} is derived
whenever employees \mytt{Em$_1$} and \mytt{Em$_2$} violate the turnover constraint for shift \mytt{Sh} and skill \mytt{Sk}.
Clearly, analogous modifications can be applied to all the problem constraints.
In this way, suitable explanation of violations can be provided so that
the user can take corrective actions.

\section{System Architecture and Usage}\label{sec:system}

\begin{figure}[t!]
\centering
\includegraphics[width=6.0cm]{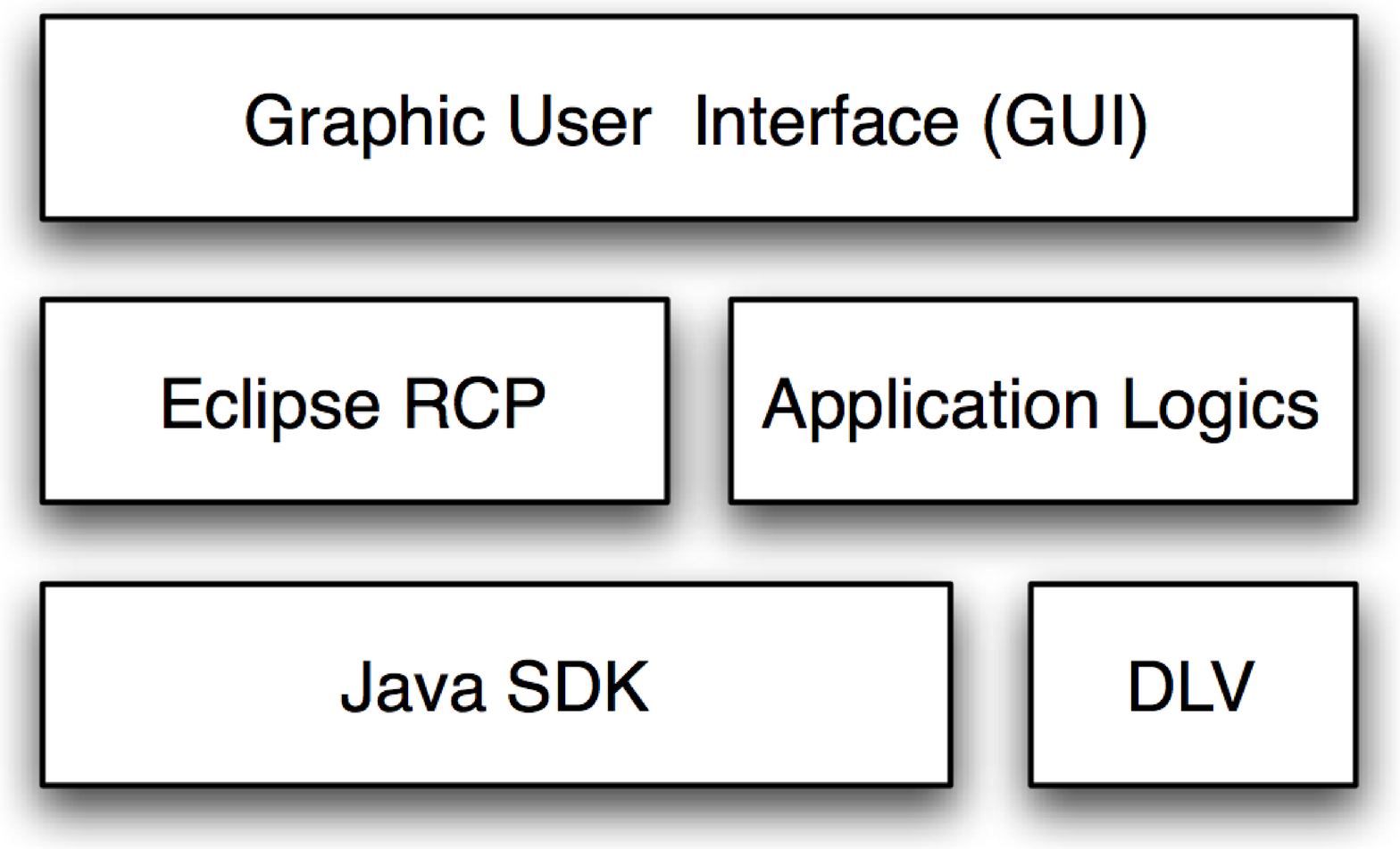}
\caption{ System Architecture}\label{fig:arch}
\end{figure}

The architecture of the team-building system is depicted in Figure~\ref{fig:arch}.
The system integrates the ASP system \dlv~\cite{leon-etal-2002-dlv} and features
a Graphical User Interface (GUI) developed in Java.
In particular, the GUI is based on the Rich Client Platform (RCP) technology
exploiting an application logics (also developed in Java)
which embeds \ontodlv~\cite{ricc-etal-2008-jlc}
(an ontology management and reasoning system based on \dlv)
for implementing both the reasoning services and data-storage features.

\begin{figure}[t!]
\centering
\includegraphics[width=18.75cm,angle=90]{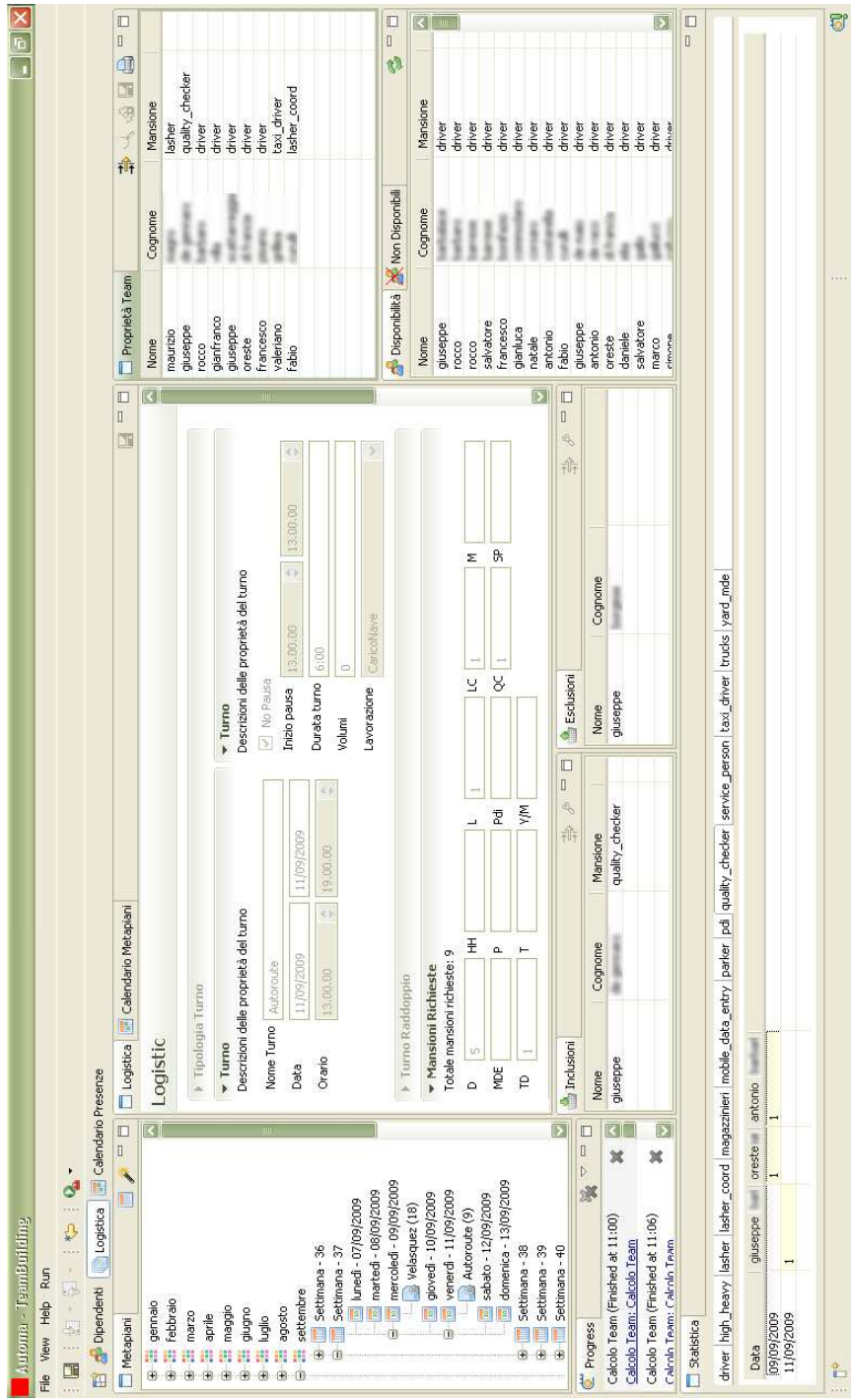}
\vspace{0.5cm}
\caption{ The Team-builder Graphic User Interface}\label{fig:gui}
\end{figure}

The GUI combines in a single customizable
frame all the  controls  (see Figure~\ref{fig:gui}). In particular,
a tree-shaped calendar (displayed on the left)
allows for browsing and scheduling working activities.
Meta-plans specifications, usually identified by the name of the
handled cargo boats (e.g., Velasquez, Autoroute),
are the leaves of the tree; they can be added or removed by right-clicking on their name
and selecting the proper command from a context-menu.
Meta-plans information (ship arrival and departure date, available processing time
and requested skills) is displayed in and editable through the ``Logistics'' panel.
Below, the ``Inclusion'' and ``Exclusion'' panels allow pre-assigning (or excluding)
specific employees from the team.
To run the team builder engine,
the user select a meta-plan (from the tree) and chooses the ``run''
item from a context-menu (activated by right-clicking the selection).%
\footnote{Note that multiple meta-plans can be executed by selecting several items from the tree.}
Once a meta-plan is run, input information and personnel statistics are fed into the \dlv system
and the result is displayed on the top-right panel (``Team Properties'').
The computed team can be also modified manually, and
the system is able to verify if the manually-modified team still satisfies
the constraints. In case of errors, causes are outlined and
suggestions for fixing a problem proposed.
The interface gives full control on the status of all the seaport-staff:
available/unavailable personnel is listed on the bottom-right panel,
and the allocation statistics are reported in the bottom panel.

\section{Some Experiments on Real-world Data}\label{sec:exp}

To provide a concrete image on the behavior of the system, we report on
a use case performed on real-world data.
In particular, we have simulated the employee-allocation on archive data provided by ICO BLG,
regarding 130 employees  and the activity of one month. We condidered four allocation activities:
($r_1$) one shift,
($r_2$) a daily allocation,
($r_3$) a weekly allocation, and
($r_4$) a monthly allocation.
The system was installed in the ICO BLG workstation featuring an
Intel Core 2 Duo CPU P8600 machine clocked at 2.40 GHz
with 4 GB of RAM running Microsoft Windows XP and Java version 1.6.0\_14.
The runtime for accomplishing the simulation of ($r_1$), ($r_2$), ($r_3$) and ($r_4$) was:
4.2 seconds, 25.3 seconds, 128.7 seconds, and 490.6 seconds, respectively.
Basically, in a few seconds the system is able to generate a shift plan,
and in some minutes one can simulate a complete allocation covering an entire month
(this feature can be used for estimating both long-range and mid-range
employees allocation needs).

In order to assess the quality of the obtained plans, we compared the results
of our system
with the manually-prepared solution provided by the ICO BLG managers.
The results confirm the higher quality of the solutions provided by our system;
indeed,
the number of constraints violations results dramatically reduced when
the system-produced allocations are compared with hand-made ones.
Further, the solutions provided by the system would have caused a decrease of
about 20\% in
the amount of overtime needed.
Those results confirmed the practical effectiveness of the ASP-based approach, and
definitely convinced the ICO BLG management to adopt our system for workforce management.

\section{Conclusion}\label{sec:conclusion}

In this paper we have presented a team building system based on Answer Set Programming.
The system features a graphical user interface that gives full control
on the status of the seaport-staff and allows for transparently running
the DLV system for computing, checking and completing suitable
teams of employees respecting the given constraints.
The system has been developed side by side with the personnel of ICO BLG
and is currently employed by the management staff of that company.
The practical effectiveness of the proposed solution
is confirmed by the on-the-field usage impressions reported by the ICO BLG
management members: ``the system is able to obtain more reliable results
when compared to manual team composition, and its use reduced
the time spent for team building by several hours each month.''

\section*{Acknowledgements}
The authors are thankful to the staff of B.L.G. at Gioia Tauro port;
a special thanks is addressed to F. Scalise, G. Lucchese, and S. Papalia.

This work has been partially supported by the Regione Calabria and EU
under POR Calabria FESR 2007-2013 within the PIA project of DLVSYSTEM.

\bibliography{./bibtex}

\newcommand{\SortNoOp}[1]{}
\begin{thebibliography}{}

\bibitem[\protect\citeauthoryear{Aickelin and Dowsland}{Aickelin and
  Dowsland}{2000}]{aick-dows}
{\sc Aickelin, U.} {\sc and} {\sc Dowsland, K.~A.} 2000.
\newblock {Exploiting problem structure in a genetic algorithm approach to a
  nurse rostering problem}.
\newblock {\em Journal of Scheduling\/}~{\em 3,\/}~3, 139--153.

\bibitem[\protect\citeauthoryear{Al-Yakoob and Sherali}{Al-Yakoob and
  Sherali}{2007}]{sale-yako-2007}
{\sc Al-Yakoob, S.} {\sc and} {\sc Sherali, H.} 2007.
\newblock {Mixed-integer programming models for an employee scheduling problem
  with multiple shifts and work locations}.
\newblock {\em Annals of Operations Research\/}~{\em 155,\/}~1, 119--142.

\bibitem[\protect\citeauthoryear{Alba and Chicano}{Alba and
  Chicano}{2007}]{alb-chic-2007}
{\sc Alba, E.} {\sc and} {\sc Chicano, J.~F.} 2007.
\newblock {Software project management with GAs}.
\newblock {\em Information Sciences\/}~{\em 177,\/}~11, 2380--2401.

\bibitem[\protect\citeauthoryear{Alfares}{Alfares}{2002}]{alfa-2002}
{\sc Alfares, H.~K.} 2002.
\newblock {Optimum workforce scheduling under the (14, 21) days-off timetable}.
\newblock {\em JAMDS\/}~{\em 6,\/}~3, 191--199.

\bibitem[\protect\citeauthoryear{Balduccini, Gelfond, Watson, and
  Nogueira}{Balduccini et~al\mbox{.}}{2001}]{bald-etal-01}
{\sc Balduccini, M.}, {\sc Gelfond, M.}, {\sc Watson, R.}, {\sc and} {\sc
  Nogueira, M.} 2001.
\newblock {The USA-Advisor: A Case Study in Answer Set Planning}.
\newblock In {\em Proceedings of LPNMR '01,}. LNCS, vol. 2173. Springer Berlin
  / Heidelberg, 439--442.

\bibitem[\protect\citeauthoryear{Baral}{Baral}{2003}]{bara-2002}
{\sc Baral, C.} 2003.
\newblock {\em {Knowledge Representation, Reasoning and Declarative Problem
  Solving}}.
\newblock Cambridge University Press.

\bibitem[\protect\citeauthoryear{Baral and Gelfond}{Baral and
  Gelfond}{2000}]{bara-gelf-2000}
{\sc Baral, C.} {\sc and} {\sc Gelfond, M.} 2000.
\newblock {\em {Logic-based artificial intelligence}}.
\newblock Kluwer Academic Publishers, Chapter {Reasoning agents in dynamic
  domains}, 257--279.

\bibitem[\protect\citeauthoryear{Baral and Uyan}{Baral and
  Uyan}{2001}]{bara-uyan-2001}
{\sc Baral, C.} {\sc and} {\sc Uyan, C.} 2001.
\newblock {Declarative Specification and Solution of Combinatorial Auctions
  Using Logic Programming}.
\newblock In {\em Logic Programming and Nonmotonic Reasoning}. LNCS, vol. 2173.
  Springer Berlin / Heidelberg, 186--199.

\bibitem[\protect\citeauthoryear{Bardadym}{Bardadym}{1996}]{bard-95}
{\sc Bardadym, V.} 1996.
\newblock {Computer-aided school and university timetabling: The new wave}.
\newblock In {\em Practice and Theory of Automated Timetabling}. LNCS, vol.
  1153. Springer Berlin / Heidelberg, 22--45.

\bibitem[\protect\citeauthoryear{Bechtold, Brusco, and Showalter}{Bechtold
  et~al\mbox{.}}{1991}]{bech-etal}
{\sc Bechtold, S.~E.}, {\sc Brusco, M.~J.}, {\sc and} {\sc Showalter, M.~J.}
  1991.
\newblock {A Comparative Evaluation of Labor Tour Scheduling Methods}.
\newblock {\em Decision Sciences\/}~{\em 22,\/}~4, 683--699.

\bibitem[\protect\citeauthoryear{Billionnet}{Billionnet}{1999}]{billi}
{\sc Billionnet, A.} 1999.
\newblock {Integer programming to schedule a hierarchical workforce with
  variable demands}.
\newblock {\em European Journal of Operational Research\/}~{\em 114,\/}~1,
  105--114.

\bibitem[\protect\citeauthoryear{Burke and Soubeiga}{Burke and
  Soubeiga}{}]{burke-soub}
{\sc Burke, E.~K.} {\sc and} {\sc Soubeiga, E.}
\newblock {A Real-World Workforce Scheduling Problem in the Hospitality
  Industry: Theoretical Models and Algorithmic Methods}.

\bibitem[\protect\citeauthoryear{Cerulli, Gaudioso, and Mautone}{Cerulli
  et~al\mbox{.}}{1992}]{gaud-ceru}
{\sc Cerulli, R.}, {\sc Gaudioso, M.}, {\sc and} {\sc Mautone, R.} 1992.
\newblock {A class of manpower scheduling problems}.
\newblock {\em Mathematical Methods of Operations Research\/}~{\em 36,\/}~1,
  93--105.

\bibitem[\protect\citeauthoryear{Chiu, Cheung, and Leung}{Chiu
  et~al\mbox{.}}{2005}]{chiu-etal}
{\sc Chiu, D. K.~W.}, {\sc Cheung, S.~C.}, {\sc and} {\sc Leung, H.-f.} 2005.
\newblock {A Multi-Agent Infrastructure for Mobile Workforce Management in a
  Service Oriented Enterprise}.
\newblock In {\em Proceedings of HICSS '05}. IEEE Computer Society, 85c--85c.

\bibitem[\protect\citeauthoryear{Dechter}{Dechter}{2004}]{dech-2004}
{\sc Dechter, R.} 2004.
\newblock {\em {Constraint processing}}.
\newblock Morgan Kaufmann Publishers.

\bibitem[\protect\citeauthoryear{Eiter, Faber, Leone, and Pfeifer}{Eiter
  et~al\mbox{.}}{2000}]{eite-etal-2000c}
{\sc Eiter, T.}, {\sc Faber, W.}, {\sc Leone, N.}, {\sc and} {\sc Pfeifer, G.}
  2000.
\newblock {\em Logic-based artificial intelligence}.
\newblock Kluwer Academic Publishers, Chapter {Declarative problem-solving
  using the DLV system}, 79--103.

\bibitem[\protect\citeauthoryear{Eitzen, Panton, and Mills}{Eitzen
  et~al\mbox{.}}{2004}]{eitz-etal}
{\sc Eitzen, G.}, {\sc Panton, D.}, {\sc and} {\sc Mills, G.} 2004.
\newblock {Multi-Skilled Workforce Optimisation}.
\newblock {\em Annals of Operations Research\/}~{\em 127,\/}~1, 359--372.

\bibitem[\protect\citeauthoryear{Ernst, Jiang, Krishnamoorthy, and Sier}{Ernst
  et~al\mbox{.}}{2004}]{erns-etal}
{\sc Ernst, A.~T.}, {\sc Jiang, H.}, {\sc Krishnamoorthy, M.}, {\sc and} {\sc
  Sier, D.} 2004.
\newblock Staff scheduling and rostering: A review of applications, methods and
  models.
\newblock {\em European Journal of Operational Research\/}~{\em 153,\/}~1,
  3--27.

\bibitem[\protect\citeauthoryear{Faber, Leone, and Pfeifer}{Faber
  et~al\mbox{.}}{2004}]{fabe-etal-2004-jelia}
{\sc Faber, W.}, {\sc Leone, N.}, {\sc and} {\sc Pfeifer, G.} 2004.
\newblock {Recursive Aggregates in Disjunctive Logic Programs: Semantics and
  Complexity}.
\newblock In {\em Logics in Artificial Intelligence}. LNCS, vol. 3229. Springer
  Berlin / Heidelberg, 200--212.

\bibitem[\protect\citeauthoryear{Franconi, Palma, Leone, Perri, and
  Scarcello}{Franconi et~al\mbox{.}}{2001}]{fran-etal-2001}
{\sc Franconi, E.}, {\sc Palma, A.}, {\sc Leone, N.}, {\sc Perri, S.}, {\sc
  and} {\sc Scarcello, F.} 2001.
\newblock {Census Data Repair: A Challenging Application of Disjunctive Logic
  Programming}.
\newblock In {\em Logic for Programming, Artificial Intelligence, and
  Reasoning}. LNCS, vol. 2250. Springer Berlin / Heidelberg, 561--578.

\bibitem[\protect\citeauthoryear{Friedrich and Ivanchenko}{Friedrich and
  Ivanchenko}{2008}]{frie-08-techrep}
{\sc Friedrich, G.} {\sc and} {\sc Ivanchenko, V.} 2008.
\newblock {Diagnosis from first principles for workflow executions}.
\newblock Tech. Rep.~2, Alpen-Adria-Universität Klagenfurt.

\bibitem[\protect\citeauthoryear{Gebser, Liu, Namasivayam, Neumann, Schaub, and
  Truszczy{\'n}skii}{Gebser
  et~al\mbox{.}}{2007}]{gebs-etal-2007-lpnmr-competition}
{\sc Gebser, M.}, {\sc Liu, L.}, {\sc Namasivayam, G.}, {\sc Neumann, A.}, {\sc
  Schaub, T.}, {\sc and} {\sc Truszczy{\'n}skii, M.} 2007.
\newblock {The First Answer Set Programming System Competition}.
\newblock In {\em Logic Programming and Nonmonotonic Reasoning}. LNCS, vol.
  4483. Springer Berlin / Heidelberg, 3--17.

\bibitem[\protect\citeauthoryear{Gelfond and Leone}{Gelfond and
  Leone}{2002}]{gelf-leon-02}
{\sc Gelfond, M.} {\sc and} {\sc Leone, N.} 2002.
\newblock {Logic programming and knowledge representation-the A-prolog
  perspective}.
\newblock {\em Artif. Intell.\/}~{\em 138,\/}~1-2, 3--38.

\bibitem[\protect\citeauthoryear{Gelfond and Lifschitz}{Gelfond and
  Lifschitz}{1991}]{gelf-lifs-91}
{\sc Gelfond, M.} {\sc and} {\sc Lifschitz, V.} 1991.
\newblock {Classical negation in logic programs and disjunctive databases}.
\newblock {\em New Generation Computing\/}~{\em 9}, 365--385.

\bibitem[\protect\citeauthoryear{Grasso, Iiritano, Leone, and Ricca}{Grasso
  et~al\mbox{.}}{2009}]{grass-etal-09-apps-lpnmr}
{\sc Grasso, G.}, {\sc Iiritano, S.}, {\sc Leone, N.}, {\sc and} {\sc Ricca,
  F.} 2009.
\newblock {Some DLV Applications for Knowledge Management}.
\newblock In {\em Logic Programming and Nonmonotonic Reasoning}. LNCS, vol.
  5753. Springer Berlin / Heidelberg, 591--597.

\bibitem[\protect\citeauthoryear{Gresh, Connors, Fasano, and Wittrock}{Gresh
  et~al\mbox{.}}{2007}]{gres-etal}
{\sc Gresh, D.~L.}, {\sc Connors, D.~P.}, {\sc Fasano, J.~P.}, {\sc and} {\sc
  Wittrock, R.~J.} 2007.
\newblock {Applying supply chain optimization techniques to workforce planning
  problems}.
\newblock {\em IBM J. Res. Dev.\/}~{\em 51,\/}~3, 251--261.

\bibitem[\protect\citeauthoryear{Hultberg and Cardoso}{Hultberg and
  Cardoso}{1997}]{hult-card-97}
{\sc Hultberg, T.~H.} {\sc and} {\sc Cardoso, D.~M.} 1997.
\newblock {The teacher assignment problem: A special case of the fixed charge
  transportation problem}.
\newblock {\em European Journal of Operational Research\/}~{\em 101,\/}~3,
  463--473.

\bibitem[\protect\citeauthoryear{Lau}{Lau}{1996}]{lau}
{\sc Lau, H.~C.} 1996.
\newblock On the complexity of manpower shift scheduling.
\newblock {\em Computers \& Operations Research\/}~{\em 23,\/}~1, 93--102.

\bibitem[\protect\citeauthoryear{Lee and Meng}{Lee and
  Meng}{2009}]{lee-meng-2009-lpnmr}
{\sc Lee, J.} {\sc and} {\sc Meng, Y.} 2009.
\newblock {On Reductive Semantics of Aggregates in Answer Set Programming}.
\newblock In {\em Logic Programming and Nonmonotonic Reasoning}. LNCS, vol.
  5753. Springer Berlin / Heidelberg, 182--195.

\bibitem[\protect\citeauthoryear{Leone, Greco, Ianni, Lio, Terracina, Eiter,
  Faber, Fink, Gottlob, Rosati, Lembo, Lenzerini, Ruzzi, Kalka, Nowicki, and
  Staniszkis}{Leone et~al\mbox{.}}{2005}]{leon-etal-2005}
{\sc Leone, N.}, {\sc Greco, G.}, {\sc Ianni, G.}, {\sc Lio, V.}, {\sc
  Terracina, G.}, {\sc Eiter, T.}, {\sc Faber, W.}, {\sc Fink, M.}, {\sc
  Gottlob, G.}, {\sc Rosati, R.}, {\sc Lembo, D.}, {\sc Lenzerini, M.}, {\sc
  Ruzzi, M.}, {\sc Kalka, E.}, {\sc Nowicki, B.}, {\sc and} {\sc Staniszkis,
  W.} 2005.
\newblock {The INFOMIX system for advanced integration of incomplete and
  inconsistent data}.
\newblock In {\em Proceedings of SIGMOD '05, Baltimore, Maryland}. ACM, New
  York, NY, USA, 915--917.

\bibitem[\protect\citeauthoryear{Leone, Pfeifer, Faber, Eiter, Gottlob, Perri,
  and Scarcello}{Leone et~al\mbox{.}}{2006}]{leon-etal-2002-dlv}
{\sc Leone, N.}, {\sc Pfeifer, G.}, {\sc Faber, W.}, {\sc Eiter, T.}, {\sc
  Gottlob, G.}, {\sc Perri, S.}, {\sc and} {\sc Scarcello, F.} 2006.
\newblock {The DLV system for knowledge representation and reasoning}.
\newblock {\em ACM Trans. Comput. Logic\/}~{\em 7,\/}~3, 499--562.

\bibitem[\protect\citeauthoryear{Lesaint, Voudouris, Azarmi, Alletson, and
  Laithwaite}{Lesaint et~al\mbox{.}}{2003}]{lesa-etal}
{\sc Lesaint, D.}, {\sc Voudouris, C.}, {\sc Azarmi, N.}, {\sc Alletson, I.},
  {\sc and} {\sc Laithwaite, B.} 2003.
\newblock {Field workforce scheduling}.
\newblock {\em BT Technology Journal\/}~{\em 21,\/}~4, 23--26.

\bibitem[\protect\citeauthoryear{Naveh, Richter, Altshuler, Gresh, and
  Connors}{Naveh et~al\mbox{.}}{2007}]{nave-etal-2007}
{\sc Naveh, Y.}, {\sc Richter, Y.}, {\sc Altshuler, Y.}, {\sc Gresh, D.~L.},
  {\sc and} {\sc Connors, D.~P.} 2007.
\newblock {Workforce optimization: Identification and assignment of
  professional workers using constraint programming}.
\newblock {\em IBM Journal of Research and Development\/}~{\em 51,\/}~3.4,
  263--279.

\bibitem[\protect\citeauthoryear{Nogueira, Balduccini, Gelfond, Watson, and
  Barry}{Nogueira et~al\mbox{.}}{2001}]{noge-etal-2001}
{\sc Nogueira, M.}, {\sc Balduccini, M.}, {\sc Gelfond, M.}, {\sc Watson, R.},
  {\sc and} {\sc Barry, M.} 2001.
\newblock {An A-Prolog Decision Support System for the Space Shuttle}.
\newblock In {\em Practical Aspects of Declarative Languages}. LNCS, vol. 1990.
  Springer Berlin / Heidelberg, 169--183.

\bibitem[\protect\citeauthoryear{Ricca, Gallucci, Schindlauer, Dell'Armi,
  Grasso, and Leone}{Ricca et~al\mbox{.}}{2009}]{ricc-etal-2008-jlc}
{\sc Ricca, F.}, {\sc Gallucci, L.}, {\sc Schindlauer, R.}, {\sc Dell'Armi,
  T.}, {\sc Grasso, G.}, {\sc and} {\sc Leone, N.} 2009.
\newblock {OntoDLV: An ASP-based System for Enterprise Ontologies}.
\newblock {\em J Logic Computation\/}~{\em 19,\/}~4, 643--670.

\bibitem[\protect\citeauthoryear{Rossi}{Rossi}{2000}]{ross-2000}
{\sc Rossi, F.} 2000.
\newblock {Constraint (Logic) Programming: A Survey on Research and
  Applications}.
\newblock In {\em New Trends in Constraints}. LNCS. Springer Berlin /
  Heidelberg, 40--74.

\bibitem[\protect\citeauthoryear{Sun, Aronson, McKeown, and Drinka}{Sun
  et~al\mbox{.}}{1998}]{ming-etal}
{\sc Sun, M.}, {\sc Aronson, J.~E.}, {\sc McKeown, P.~G.}, {\sc and} {\sc
  Drinka, D.} 1998.
\newblock {A tabu search heuristic procedure for the fixed charge
  transportation problem}.
\newblock {\em European Journal of Operational Research\/}~{\em 106,\/}~2-3,
  441--456.

\bibitem[\protect\citeauthoryear{Tien and Kamiyama}{Tien and
  Kamiyama}{1982}]{tien-etal}
{\sc Tien, J.~M.} {\sc and} {\sc Kamiyama, A.} 1982.
\newblock On manpower scheduling algorithms.
\newblock {\em SIAM Review\/}~{\em 24,\/}~3, 275--287.

\bibitem[\protect\citeauthoryear{Vacca, Bierlaire, and Salani}{Vacca
  et~al\mbox{.}}{2007}]{vacca-etal-2007}
{\sc Vacca, I.}, {\sc Bierlaire, M.}, {\sc and} {\sc Salani, M.} 2007.
\newblock {Optimization at Container Terminals: Status, Trends and
  Perspectives}.
\newblock Swiss Transport Research Conference, Ascona, Switzerland, September
  14, 2007.

\bibitem[\protect\citeauthoryear{Wren and Wren}{Wren and Wren}{1995}]{wren}
{\sc Wren, A.} {\sc and} {\sc Wren, D.~O.} 1995.
\newblock A genetic algorithm for public transport driver scheduling.
\newblock {\em Computers \& Operations Research\/}~{\em 22,\/}~1, 101--110.

\bibitem[\protect\citeauthoryear{Yang}{Yang}{1996}]{yang}
{\sc Yang, R.} 1996.
\newblock {Solving a Workforce Management Problem with Constraint Programming}.
\newblock In {\em Proceedings of PACT '96}. The Practical Application Company
  Ltd, 373--387.

\end{thebibliography}

\end{document}